\documentclass[aps,pre,epsfig,floats,twocolumn,superscriptaddress,amssymb,amsmath,floatfix,showpacs]{revtex4}
\usepackage{graphicx}
\usepackage{bm}  % bold math
\usepackage{amssymb}
\usepackage{color}
\usepackage{amscd}

\begin{document}
\title{Two-sphere low Reynold's propeller}

\author{Ali Najafi}
\email{najafi@znu.ac.ir} \affiliation{Department of Physics,
Zanjan University, Zanjan 313, Iran}

\author{Rojman Zargar}
\affiliation{Institute for Advanced Studies in Basic Sciences
(IASBS), P. O. Box 45195-1159, Zanjan 45195, Iran}

\date{\today}

\begin{abstract}
A three-dimensional model of a low-Reynold's swimmer is introduced and analyzed in this paper. This model consists of two large and small 
spheres connected by two perpendicular thin rods. The geometry of this system is motivated by the microorganisms that use a single tail 
to swim, the large sphere represents the head of microorganism and the small sphere resembles its tail. Each rod changes its length 
and orientation in a non-reciprocal manner that effectively propel the system.  
%The numerical solutions are presented for the dynamical equations of this system. 
Translational and rotational velocities of the swimmer are studied for different values 
of parameters. Our findings show that by changing the parameters we can adjust both the velocity and the direction of motion of the 
swimmer. %Our modeled swimmer suggests a proposal for  human-made micro-machines and also a simplified theoretical model to 
%study the  dynamical behavior of microorganisms. 
\end{abstract}

%  87.19.ru   locomotion
%  47.15.G-   Low-Reynolds-number (creeping) flows
%  67.40.Hf   Hydrodynamics in specific geometries, flow in narrow channels
%  45.40.Ln   Robotics

\maketitle
\section{Introduction}
The propulsive motion of artificial and biological micron-scale objects is an interesting problem at low Reynolds hydrodynamics. In this 
condition the dynamics is dominated by viscous forces.
Examples of these micron scale objects include biological microorganisms like bacteria and also man-made  micro-swimmers, useful to operate at microfluidic investigations \cite{micro}. 

Propulsive motion at low Reynold's number is subject to the \emph{Scallop theorem} \cite{Purcell}. At small scales, where, the Reynolds number is very low, the governing hydrodynamic equations, i.e. the Stokes 
and continuity equations are linear and invariant under time reversal \cite{Happel}. Any reciprocal shape deformation retraces its trajectory and the system 
stay back at the point where it started. In order to achieve a net translational displacement, the system should perform the body deformations in a non-reciprocal 
manner. As mentioned by Purcell a low-Reynolds propeller must have at least two 
internal degrees of freedom and he proposed a three-link swimmer. The detailed motion of Purcell's swimmer was examined by H. Stone where, 
it was shown that  Purcell's system could swim and its dynamical properties were calculated \cite{Stone}. Inspired by Purcell's system, a  low 
Reynolds swimmer constructed by three linked spheres was introduced and analyzed by Najafi {\it et al.} \cite{Najafi}, and experimentally realized by M. Leoni {\it et al.} \cite{3sexp}.
After Purcell's proposal there have been considerable scientific efforts in designing artificial swimmers. 
Such swimmers would be useful in developing microfluidic experiments. Furthermore, progresses in assembling micro-swimmers 
show the possibility of using micro-machines inside the biological cells for non invasive therapeutic treatments \cite{Ishiyama}. 
On the other hand,  there are many theoretical works devoted to the study of different aspects in the motion of biological 
microorganisms at low Reynolds number condition \cite{new1,new2,new3,new4,new5}. Such interests include sperm swimming, metachronal waves in cilia, E-Coli chemotaxis and 
coupling mediated by hydrodynamic interaction  between nearby microorganisms \cite{Ecoli, Frank}.
For a review of recent progress on low Reynolds hydrodynamics of microorganisms, see for example the review article by E. Lauga \cite{Lauga}.
\begin{figure}
  \centerline{\includegraphics[width=.80\columnwidth]{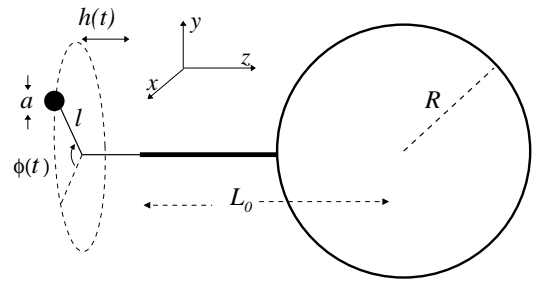}}% Images in 100% size
  \caption{Schematic showing the geometry of a two-sphere swimmer. Two large and small spheres are connected through 
two perpendicular rods, one with fixed but the other with variable length. The short rod is rotating around the long rod. 
This model system resembles the motion of a bacterium that has a single tail.}\label{fig1}
\end{figure}

Our first aim in this article is to present a simplified model that captures the characteristics of a swimming biological organism like a bacterium. Dipolar far velocity field and asymmetric shape, corresponding to the head-tail geometry of the organisms are two important 
features of micro swimmers. We model these systems by considering two spheres with different radii that are changing their separation. We will 
study the translational and angular motion of this system.
%
%In this paper we focus on a three dimensional model system constructed by two connected large and small spheres. 

%The rest of this paper is organized as below. In section 2, we 
%introduce the detail geometry of the two-sphere system. Mathematical details of a point-force (Stokeslett) in the presence of a rigid 
%no-slip sphere is reviewed and outlined in section 3. Section 4, presents the mathematical equations governing the dynamics of two-sphere 
%system. Numerical solutions to these governing equations and also discussion about our results are presented in section 5.

\section{Two-sphere model}
Figure \ref{fig1} shows the schematic geometry of a model swimmer composed of two spheres. As shown in this figure 
two small and large spheres with radii  $a$ and $R$ are connected by two perpendicular and negligible diameter rods. Let denote the 
lengths of long and short rods by $L$ and $l$ respectively. The connection is established in a way that the angle between two rods is fixed to $\frac{\pi}{2}$ while the relative angular position of small rod with respect to the large sphere can be varied. Additionally, we assume that the length of the long 
rod can be dynamically changed. In this case, the system will have two internal degrees of freedom, the length of the long rod $L(t)$ and 
the rotational angle of the short rod $\phi(t)$. 

The geometry which we are introducing here resembles the body shape of a bacterium with a single flagellum or cilium. Bacteria use beating patterns in their tails to move. The small sphere in our two-sphere model acts as a beating tail and  the large sphere resembles the 
head of animal. 
The minimum condition for swimming at low Reynolds number can be achieved in our three-dimensional model. By changing the length of long rod and the angle of small one in a prescribed 
form we are able to choose the motion which breaks the time-reversal symmetry, the necessary condition for translational motion and consequently propel the system.
\begin{figure}
  \centerline{\includegraphics[width=.80\columnwidth]{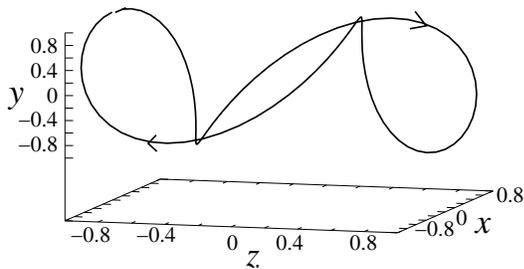}}% Images in 100% size
  \caption{Trajectory of the small sphere seen in the frame of reference that is co-moving with large sphere. Here we chose $h_0=1$ 
and $\frac{\omega_{\phi}}{\omega_L}=2$.}\label{fig2}
\end{figure}

As an example for the internal motion of the system, we let the angle $\phi(t)$ increase with constant angular velocity and the length of long rod change periodically around an average length. The explicit form of this motion is given by: 
$L(t)=L_0+h_0\cos(\omega_{L}t-\phi_0)$, and $\phi(t)=\omega_{\phi}t$.
%\begin{equation}
%L(t)=L_0+h_0\cos(\omega_{L}t-\phi_0),~~\phi(t)=\omega_{\phi}t
%\nonumber.
%\end{equation}
In this case, the position vector of the small sphere, seen in the reference frame that is co-moving with the 
large sphere, is given by:
\begin{equation}
{\bf X}_0=(l \cos\phi(t),l \sin\phi(t),L_0+h(t)),
\label{21}
\end{equation}
where $l$ is the length of the small rod and $L_0$ represent  the average length of the long rod.
Figure \ref{fig2} shows a typical real space trajectory of the small sphere that is seen in the reference frame co-moving with the large sphere.
Different choices for $\omega_{L}$ and $\omega_{\phi}$ correspond to different forms of the internal motion. 
For $\frac{\omega_{\phi}}{\omega_{L}}=m=\frac{p}{q}$ with $p$ and $q$ as two positive integer numbers, we see that the paths in the 
$(h, \phi)$-space are closed loops. One should note that the phase space of internal motion 
($(h, \phi)$-space) is the surface of a cylinder. The axial direction on the cylinder represents the $h$-direction and the azimuthal angle is shown by the transverse direction on the cylinder.  
For $m<1$, the phase space trajectory is a closed curve which traces exactly one complete turn round the cylinder while for $m\ge 1$ the 
trajectory is a closed loop that  turns many times 
round the cylinder. In both cases, the geometry of the closed curves in the cylindrical-shape phase space are examples of nonreciprocal 
motions that can generate a net translational motion.
\section{Point Force Near A Rigid Sphere}
%Here we review the hydrodynamics for low Reynolds condition.
%Here we are dealing with the dynamics of an immersed object in an incompressible viscous fluid with prescribed geometrical structure.
At zero Reynold's number the Stokes equations govern the dynamics of fluid. 
%Denoting the velocity and pressure fields of the fluid by 
%${\bf u}({\bf X})$ and $p({\bf X})$ the governing equations are:
%\begin{equation}
%\eta\nabla^2{\bf u}({\bf X})=\nabla p({\bf X}),~~~~~~~~~~~~~~~~~\nabla\cdot{\bf u}({\bf X})=0,
%\end{equation}
%where $\eta$ is the viscosity of the fluid. Here the fluid is assumed to be incompressible. The fluid velocity field is subject to no-slip %boundary condition on the surface of solid objects. 
%
The solution of the Stokes equation for a point force singularity is formulated in terms of Green's function and is called Stokeslet. For a point force singularity with strength ${\bf f}$ located at point ${\bf X}_{0}$, the velocity field generated in the fluid is given by: 
${\bf u}({\bf X})=\frac{1}{8\pi\eta}  G({\bf X},{\bf X}_0)\cdot {\bf f}$,
%\begin{equation}
%{\bf u}({\bf X})=\frac{1}{8\pi\eta}  G({\bf X},{\bf X}_0)\cdot {\bf f},
%\end{equation}
where $\eta$ is the viscosity of the fluid. C.W. Oseen has derived the explicit form of the Green's function $G$ for an infinite flow that is bounded internally by a 
solid sphere with radius $R$ \cite{Oseen}. For the explicit form of this solution, we refer to the original article by Oseen. 
As it is manifested by Oseen's solution, the flow field due to a point force in the presence of a solid sphere can be 
regarded as the flow of the original point force and the flow due to the  singular parts that are located at an image position 
inside the sphere. The location of the image 
point inside the sphere is given by its position vector, ${\bf X}_0^{*}=\frac{R^2}{X_0^2}{\bf X}_0$, relative to the sphere's center.

As argued by Higdon the total force acting by a point force on fluid bounded by a 
no-slip sphere is equivalent to the total Stokeslet strength \cite{Hig}. Total Stokeslet strength includes the image point force inside the sphere. For a 
point force ${\bf f}$, the image point force is defined by: ${\bf f}_{I}=(c_{r}{\bf f}_{r}+c_{t}{\bf f}_{t})$
where the radial and tangential components of this image force are given by: $
{\bf f}_r={\bf f}\cdot{\bf X}_0/X_0,~~~~~{\bf f}_t={\bf f}-{\bf f}_r$. Here two coefficients $c_r$ and $c_t$ are given by:
\begin{equation}
c_r=-\frac{3}{2}\frac{R}{X_{0}}+\frac{1}{2}\frac{R^{3}}{X_{0}^{3}} ~~,~~
c_t=-\frac{3}{4}\frac{R}{X_{0}}-\frac{1}{4}\frac{R^{3}}{X_{0}^{3}}.
\end{equation}
In the next section we will use these results for analyzing the motion of two-sphere system.
\section{Two-Sphere Dynamics}
In this section we will develop the dynamical equations for two-sphere system. To simplify the equations We will assume 
that the radius of small sphere is much smaller than the radius of large sphere ($a\ll R$). We further assume that $a$  is smaller than the characteristic distance between the spheres. With this approximation the velocity field of the system is described by the velocity field of a point force that is moving near a rigid sphere. This simplification allows us to use the results of the preceding section and derive a more 
simpler dynamical equations of the system. 
However one should note that the finite size effect of the small sphere can be systematically considered by Faxen's theorem \cite{Kim}. 
%As shown 
%by Y.W. Kim, the mobility tensor of a spherical object with radius $a$ is related to the point force Green's function by \cite{Kim}:
%\begin{equation}
%{\bf \mu}({\bf x}_i,{\bf x}_j)=\left(1+\frac{a^2}{6}\nabla^{2}_{{\bf x}_i}\right)
%\left(1+\frac{a^2}{6}\nabla^{2}_{{\bf x}_j}\right)
% G({\bf x}_i,{\bf x}_j).\nonumber
%\end{equation}
%This formula will allow us to obtain the size effects of small sphere by replacing the point force Green's function with the above 
%mobility tensor.

To obtain the swimming velocity of the system we work in the reference frame, that is co-moving and rotating with the large sphere. In this 
coordinate system the velocity of the fluid at infinity is the swimming velocity. Denoting the swimming velocity of the system by $V$, and its 
angular velocity by $\Omega$, we can express the velocity field of the fluid at a general point ${\bf X}$ as:
\begin{equation}
{\bf u}({\bf X})=-{\bf V}-{\bf \Omega}\times {\bf X}+ M.{\bf V}+{\bf \Omega}\times {\bf m}+ G.{\bf f},\label{42}
\end{equation}
where the tensor $ M$ and vector ${\bf m}$ give the flow field due to the translational and rotational motion of a moving sphere. These 
quantities are given by:
\begin{equation}
 M=\frac{3}{4}\frac{R}{X}[{\bf I}+\frac{{\bf X}{\bf X}}{X^{2}}]+\frac{1}{4}\frac{R^{3}}{X^{3}}[{\bf I}-3\frac{{\bf X}{\bf X}}{X^{2}}],~~~~
{\bf m}=\frac{R^{3}}{X^{3}} {\bf X} \label{43}
\end{equation}

In the absence of external force and torque, the swimmer is force and torque free. Therefore we require that the total force and 
torque acting on the fluid by the system to be zero. Including the point force and its image counter part and adding the contributions due to the translational and rotational 
motion of large sphere we arrive at the following force and torque balance equations:
%\begin{eqnarray}
%&~&{\bf f}+{\bf f}_{I}+6\pi\eta R {\bf V}=0 \nonumber \\
%&~&{\bf X}_{0} \times {\bf f}_{t}-\frac{R^{3}}{X_{0}^{3}}{\bf X}_{0} \times {\bf f}_{t}+8\pi\eta R^{3}{\bf \Omega}=0
%\label{44}
%\end{eqnarray}
\begin{eqnarray}
\left\{
\begin{array}{l}
 {\bf f}+{\bf f}_{I}+6\pi\eta R {\bf V}=0 \\
 {\bf X}_{0} \times {\bf f}_{t}-\frac{R^{3}}{X_{0}^{3}}{\bf X}_{0} \times {\bf f}_{t}+8\pi\eta R^{3}{\bf \Omega}=0
\end{array}
\right.
                   \label{44}
\end{eqnarray}
%In this approach starting with an initial condition we should be able to extract the swimming velocity as well as the rotational velocity 
%of the system.
%Our aim is to calculate the swimming velocity for a system with some prescribed internal motion. 
The fluid velocity field at the location of small sphere is subject to the boundary condition: ${\bf u}({\bf X}_0)=\dot{{\bf X}}_0$. 
Together with this boundary condition, the above equations make a complete set of dynamical governing equations for the swimmer.

To solve the dynamical equations for the system, we can use the force and torque balance conditions and obtain a set of equations 
which relate the different components of the translational or angular velocities of the system to the component of the vector $\dot {{\bf X}}_0$ in the following matrix form:
\begin{eqnarray}
{\bf V}&=&{\bf A}~~{\bf C}^{-1}~~\dot {{\bf X}}_0\nonumber\\
{\bf \Omega}&=&{\bf B}~~{\bf C}^{-1}~~\dot {{\bf X}}_0
\label{46}
\end{eqnarray}
where the detail for of the matrix elements $[A]_{ij}=a_{ij}$, $[B]_{ij}=b_{ij}$ and $[C]_{ij}=c_{ij}$ are given at the appendix.
%\begin{eqnarray}
%&&V_{x}=a_{11}f_{x}+a_{12}f_{y}+a_{13}f_{z} \nonumber \\
%&&V_{y}=a_{21}f_{x}+a_{22}f_{y}+a_{23}f_{z}  \nonumber\\
%&&V_{z}=a_{31}f_{x}+a_{32}f_{y}+a_{33}f_{z}.  
%\label{46}
%\end{eqnarray}
\begin{figure}
  \centerline{\includegraphics[width=.80\columnwidth]{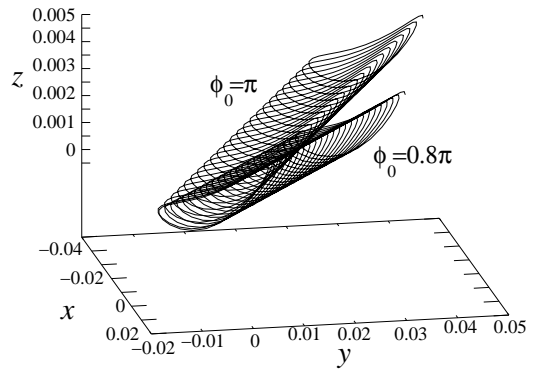}}% Images in 100% size
  \caption{The trajectory of the swimmer in the ($x$, $y$, $z$)-space is plotted for two different values of $\phi_0$. 
Other parameters are $R=1$, $a=0.5$, $L_0=4$, $h_0=0.1$, $l=0.3$ and $\omega_{\phi}=2\omega_l=1$. Line show the real path of the large sphere. The swimmer starts its motion from the initial state where the large sphere is located at the origin and the long rod is orientated along 
the $-\hat{z}$ direction. As one can see, the overall swimming direction can be varied by changing the parameters of the system.
Average orientation of the long rod which is not shown here is along  the average 
swimming direction.}\label{fig3}
\end{figure}
\begin{figure}
  \centerline{\includegraphics[width=.80\columnwidth]{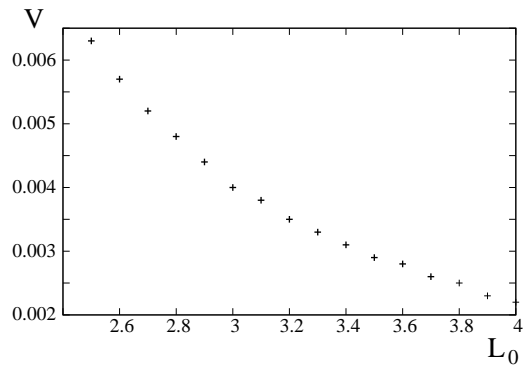}}% Images in 100% size
  \caption{Average swimming velocity is plotted in terms of the length of long rod. 
Other parameters are set to: $R=1$, $a=0.5$, $\phi_0=0$, $h_0=0.1$, $l=0.3$ and $\omega_{\phi}=\omega_l=1$.}\label{fig4}
\end{figure}
%\begin{eqnarray}
%&&\Omega_{x}=b_{11}f_{x}+b_{12}f_{y}+b_{13}f_{z} \nonumber \\
%&&\Omega_{y}=b_{21}f_{x}+b_{22}f_{y}+b_{23}f_{z} \nonumber \\
%&&\Omega_{z}=b_{31}f_{x}+b_{32}f_{y}+b_{33}f_{z}, 
%\label{47}
%\end{eqnarray}
%where the detail form of the coefficients $a_{ij}$ and $b_{ij}$ are given at the appendix.
%We now proceed to calculate the strength of the point force. 
%Applying the boundary condition (\ref{45}) and eliminating both the linear and angular velocities of the large sphere and using 4.6 and 4.7 %we can reach the following equations:
%\begin{equation}
%{\bf f}={\bf C}^{-1}~\cdot~{\bf X}_0
%\label{48}
%\end{equation}
%Applying the condition that the velocity field of the fluid in the position 
%of point force is the prescribed velocity given by $\dot {{\bf X}}_0$ we can eliminate both linear and angular velocity of the large 
%sphere and obtain the following equations:
%\begin{eqnarray}
%&&\dot {x}_{0}=c_{11}f_{x}+c_{12}f_{y}+c_{13}f_{z} \nonumber \\
%&&\dot {y}_{0}=c_{21}f_{x}+c_{22}f_{y}+c_{23}f_{z} \nonumber \\
%&&\dot{z}_{0}=c_{31}f_{x}+c_{32}f_{y}+c_{33}f_{z} 
%\label{48}
%\end{eqnarray}
%where the coefficients $c_{ij}$ are given at the appendix. These equations relate the strength of the point force to the prescribed motion 
%of the small sphere. By solving these equations we can calculate the point force components. Having in hand the point force strength we can 
%use the set of equations (\ref{46}) and (\ref{47}) and derive the velocity and correspondingly the trajectory of large sphere in space. In %section 5 we will solve these equations.
\section{Results And Discussion}
In this section we will present the  numerical solution for the governing equations and obtain the real space trajectory of the swimmer. For this purpose we  plot the trajectory of large sphere. 
For the prescribed internal motion given by (\ref{21}) and a special choice of parameters, we have plotted figure \ref{fig3}, 
the space trajectory of the large sphere. As one can distinguish, the trajectory is a helical-shaped path with overall 
translational movement in each round. The different characteristics of the trajectory, preferred direction, average swimming velocity and the 
effective radius of the helice can be controlled by the geometrical as well as dynamical parameters of the swimmer. 

The average orientation of the long rod, which is not shown in figure, is in the direction of the longitudinal axis of helix. This is achieved by 
numerically solving for the rotational velocity. Controlling and adjusting the dynamical behavior of the swimmer is of prime importance in designing artificial micro-machines. 
Here we see that by changing the parameters of the system we can do this favor. In figure \ref{fig3}, we have shown that the overall swimming direction is  sensitive to the initial phase $\phi_0$. 
Additionally and as an another example, in figure \ref{fig4}, we have shown that by changing $L_0$, the length of long rod, the average swimming velocity can be changed.

As the geometry of the two-sphere swimmer is not symmetric, 
the far field distribution of fluid velocity at the leading order of approximation, resembles a velocity field of a single force dipole. This is the main characteristic of most swimming microorganisms with, dipole-like velocity pattern.

In summary, inspired by bacterium swimming, we proposed and analyzed a swimmer, constructed by two joint spheres. We have shown that 
this simple three-dimensional swimmer is a model for a low Reynolds propeller. %The two-sphere swimmer model
captures a number of dynamical features in microorganisms. 
It will be interesting to use this model swimmer and 
study many interesting problems like, the hydrodynamic interaction between such swimmers, the effects due to the  confinement in the bounded fluids and also chemotaxis phenomena. Inspired by the colonies of microorganisms, we are extending our model to investigate the hydrodynamic interaction in an ensemble of two-sphere swimmers.

\appendix
\section{Mathematical Details}
Here we present the explicit form of the matrix elements $a_{ij}$, $b_{ij}$ and $c_{ij}$ which  were introduced in the text:
\begin{eqnarray}
&&a_{ii}=-\frac{1}{6\pi \eta R}[(1+c_{t})+(c_{r}-c_{t})(\frac{X_{0i}^{2}}{X_{0}^{2}})] \nonumber \\
&&a_{ij}=a_{ji}=-\frac{1}{6\pi \eta R}(c_{r}-c_{t})(\frac{X_{0i}X_{0j}}{X_{0}^{2}}) ~~~for~~~i\neq j \nonumber\\
&&b_{ii}=0 \nonumber \\
&&b_{ij}=-b_{ji}=\frac{1}{8\pi \eta R^{3}}(1-\frac{R^{3}}{X_{0}^{3}})X_{0k} ~~~for~~~i\neq j\neq k \nonumber
%&&a_{11}=-\frac{1}{6\pi \eta R}[(1+c_{t})+(c_{r}-c_{t})(\frac{x_{0}^{2}}{X_{0}^{2}})] \nonumber \\
%&&a_{22}=-\frac{1}{6\pi \eta R}[(1+c_{t})+(c_{r}-c_{t})(\frac{y_{0}^{2}}{X_{0}^{2}})] \nonumber \\
%&&a_{33}=-\frac{1}{6\pi \eta R}[(1+c_{t})+(c_{r}-c_{t})(\frac{z_{0}^{2}}{X_{0}^{2}})] \nonumber \\
%&&a_{12}=a_{21}=-\frac{1}{6\pi \eta R}(c_{r}-c_{t})(\frac{x_{0}y_{0}}{X_{0}^{2}}) \nonumber \\
%&&a_{13}=a_{31}=-\frac{1}{6\pi \eta R}(c_{r}-c_{t})(\frac{x_{0}z_{0}}{X_{0}^{2}}) \nonumber \\
%&&a_{23}=a_{32}=-\frac{1}{6\pi \eta R}(c_{r}-c_{t})(\frac{y_{0}z_{0}}{X_{0}^{2}}) \nonumber
\end{eqnarray}
%\begin{eqnarray}
%&&b_{ii}=0 \nonumber \\
%&&b_{ij}=-b_{ji}=\frac{1}{8\pi \eta R^{3}}(1-\frac{R^{3}}{X_{0}^{3}})X_{0k} ~~~for~~~i\neq j\neq k \nonumber
%&&b_{11}=0 \nonumber \\
%&&b_{22}=0 \nonumber \\
%&&b_{33}=0 \nonumber \\
%&&b_{12}=-b_{21}=\frac{1}{8\pi \eta R^{3}}(1-\frac{R^{3}}{X_{0}^{3}})z_{0} \nonumber \\
%&&b_{13}=-b_{31}=\frac{1}{8\pi \eta R^{3}}(1-\frac{R^{3}}{X_{0}^{3}})y_{0} \nonumber \\
%&&b_{23}=-b_{32}=\frac{1}{8\pi \eta R^{3}}(1-\frac{R^{3}}{X_{0}^{3}})x_{0} \nonumber
%\end{eqnarray}
\begin{eqnarray}
c_{11}&=&(M_{xx}-1)a_{11}+M_{xy}a_{21}+M_{xz}a_{31}+(m_{z}-z_{0})b_{21}\nonumber \\
&&-(m_{y}-y_{0})b_{31}+G_{xx} \nonumber \\
c_{12}&=&(M_{xx}-1)a_{12}+M_{xy}a_{22}+M_{xz}a_{32}-(m_{y}-y_{0})b_{32}\nonumber \\
&&+G_{xy} \nonumber \\
c_{13}&=&(M_{xx}-1)a_{13}+M_{xy}a_{23}+M_{xz}a_{33}+(m_{z}-z_{0})b_{23}\nonumber \\
&&+G_{xz} \nonumber \\
c_{21}&=&M_{yx}a_{11}+(M_{yy}-1)a_{21}+M_{yz}a_{31}+(m_{x}-x_{0})b_{31}\nonumber \\
&&+G_{yx} \nonumber \\
c_{22}&=&M_{yx}a_{12}+(M_{yy}-1)a_{22}+M_{yz}a_{32}+(m_{x}-x_{0})b_{32}\nonumber \\
&&-(m_{z}-z_{0})b_{12}+G_{yy} \nonumber \\
c_{23}&=&M_{yx}a_{13}+(M_{yy}-1)a_{23}+M_{yz}a_{33}-(m_{z}-z_{0})b_{13}\nonumber \\
&&+G_{yz} \nonumber \\
c_{31}&=&M_{zx}a_{11}+M_{zy}a_{21}+(M_{zz}-1)a_{31}-(m_{x}-x_{0})b_{21}\nonumber \\
&&+G_{zx} \nonumber \\
c_{32}&=&M_{zx}a_{12}+M_{zy}a_{22}+(M_{zz}-1)a_{32}+(m_{y}-y_{0})b_{12}\nonumber \\
&&+G_{zy} \nonumber \\
c_{33}&=&M_{zx}a_{13}+M_{zy}a_{23}+(M_{zz}-1)a_{33}+(m_{y}-y_{0})b_{13}\nonumber \\
&&-(m_{x}-x_{0})b_{23}+G_{zz} \nonumber
\end{eqnarray}

\end{document}